\documentclass[10pt,letterpaper]{article}
\usepackage[top=0.85in,left=2.75in,footskip=0.75in]{geometry}

\usepackage{amsmath,amssymb}

\usepackage{changepage}

\usepackage[utf8x]{inputenc}

\usepackage{textcomp,marvosym}

\usepackage{cite}

\usepackage{nameref,hyperref}

\usepackage[right]{lineno}

\usepackage{microtype}
\DisableLigatures[f]{encoding = *, family = * }

\usepackage[table]{xcolor}

\usepackage{array}

\usepackage{longtable}
\newcolumntype{+}{!{\vrule width 2pt}}

\newlength\savedwidth

\raggedright
\usepackage[aboveskip=1pt,labelfont=bf,labelsep=period,justification=raggedright,singlelinecheck=off]{caption}
\renewcommand{\figurename}{Fig}

\makeatletter
\renewcommand{\@biblabel}[1]{\quad#1.}
\makeatother

\usepackage{lastpage,fancyhdr,graphicx}
\usepackage{epstopdf}
\usepackage{soul}
\usepackage{rotating}
\usepackage{circuitikz}
\fancyheadoffset[L]{2.25in}
\fancyfootoffset[L]{2.25in}
\usepackage{comment}
\excludecomment{removedforplos} %
\hypersetup{final}

\begin{document}
\vspace*{0.2in}

\begin{flushleft}
{\Large
\textbf{Inhibitory feedback from the motor circuit gates mechanosensory processing in \textit{C. elegans}}} 
\newline

Sandeep Kumar\textsuperscript{1},
Anuj K Sharma\textsuperscript{2},
Andrew Tran\textsuperscript{1},
Andrew M Leifer\textsuperscript{1,2*}\\
\bigskip
\textbf{1} Princeton Neuroscience Institute , Princeton University, Princeton, NJ, United States of America
\\
\textbf{2} Department of Physics, Princeton University, Princeton, NJ, United States of America
\\
\bigskip

* leifer@princeton.edu

\end{flushleft}
\section*{Abstract}
Animals must integrate sensory cues with their current behavioral context to generate a suitable response. How this integration occurs is poorly understood.
Previously we developed high throughput methods to probe neural activity in populations of \textit{Caenorhabditis elegans} and discovered that the animal's mechanosensory processing is rapidly modulated by the animal's locomotion. Specifically we found that when the worm turns it suppresses its mechanosensory-evoked reversal response. Here we report that \textit{C. elegans} use inhibitory feedback from turning-associated neurons to provide this rapid modulation of mechanosensory processing. By performing high-throughput optogenetic perturbations triggered on behavior, we show that turning associated neurons SAA, RIV and/or SMB suppress mechanosensory-evoked reversals during turns. We find that activation of the gentle-touch mechanosensory neurons or of any of the interneurons AIZ, RIM, AIB and AVE during a turn is less likely to evoke a reversal than activation during forward movement. Inhibiting neurons SAA, RIV and SMB during a turn restores the likelihood with which mechanosensory activation evokes reversals.  Separately,  activation of premotor interneuron AVA evokes reversals regardless of whether the animal is turning or moving forward. We therefore propose that inhibitory signals from SAA, RIV and/or SMB  gate mechanosensory signals upstream of neuron AVA. We conclude that \textit{C. elegans} rely on inhibitory feedback from the motor circuit to modulate its response to sensory stimuli on fast timescales. This need for motor signals in sensory processing may explain the ubiquity in many organisms of motor-related neural activity patterns seen across the brain, including in sensory processing areas.

\section*{Introduction}
A critical role of the nervous system is to detect sensory information  and select a suitable motor response, taking into consideration the animal's environment and current behavior.  How the brain integrates sensory stimuli with broader context is an active area of research.  For example, primates integrate a primary  visual cue with a contextual visual cue to flexibly alter their neural computations \cite{mante_context-dependent_2013, remington_flexible_2018}. In \textit{Drosophila}, dopaminergic signals reflect mating drive, a long-lived internal state, that in turn  gates the animal's courtship response to auditory and visual cues \cite{zhang_dopaminergic_2016}. In \textit{C. elegans} long-lived internal states lasting many minutes such as hunger \cite{ghosh_neural_2016}, quiescence \cite{raizen_lethargus_2008, schwarz_reduced_2011, nagy_homeostasis_2014, nagy_measurements_2014, cho_-chip_2018} and arousal \cite{cho_multilevel_2014} are all thought to alter the animal's response to stimuli via various synaptic or  neuromodulatory mechanisms  and have been shown to modulate mechanosensory response \cite{chen_modulation_2014, chen_regulation_2015}. In those investigations, sensory signals are combined with one another or are integrated with  long-lived internal state. Less is known about how sensory processing is modulated  by  short-timescale  behavior. Short seconds-timescale modulation of sensory processing is of particular  interest because  1) it allows the animal to respond to urgent signals, such as threats and 2) because the timescale suggests a circuit level mechanism, instead of other longer timescale mechanisms, such as neuromodulation or changes in gene expression.  
Here we investigate  short-timescale behavioral modulation of the \textit{C. elegans} gentle-touch response.

We study the nematode \textit{C. elegans} because its compact brain is well suited for  investigations spanning sensory input to motor output \cite{clark_mapping_2013,calhoun_quantifying_2017}. The \textit{C. elegans} gentle-touch circuitry  allows the animal to  avoid predation and is one of the most well-studied circuits of the worm \cite{chalfie_developmental_1981, chalfie_neural_1985, maguire_c_2011}. We previously discovered that animals  traveling forward are much more likely to respond to a mechanosensory stimulus by backing up (reversal), than animals that receive the same stimulus while they are in the middle of a turn  \cite{liu_temporal_2018, liu_high-throughput_2022}. In other words the worm's response to mechanosensory stimuli is gated by  the animal's  short-timescale behavioral context.  Suppressing mechanosensory-evoked reversals during turns may be part of a predator avoidance strategy. Turns are an important part of the \textit{C. elegans} escape response, and by  preventing turns from being interrupted prematurely, the animal may be ensuring that the escape response continues to completion  \cite{pirri_neuroethology_2012, liu_temporal_2018, wang_flexible_2020}.  

The neural mechanism underlying this rapid modulation of sensorimotor processing has not previously been described. Because turns are short-lived, lasting less than 2 seconds, we suspect  gating is mediated by fast neural dynamics at the circuit level. 

In mouse, fly and \textit{C. elegans},  regions across the brain  exhibit activity patterns related to the animal's locomotory state and body pose \cite{hallinen_decoding_2021, stringer_spontaneous_2019, musall_single-trial_2019, atanas_brain-wide_2022}. A leading hypothesis is that  these motor signals may be important  to  modulate sensory representations including but no limited to vision \cite{niell_modulation_2010}, thermosensation \cite{ji_corollary_2021}, or  corollary discharge \cite{crapse_corollary_2008, riedl_tyraminergic_2022}. In this study, we sought to investigate how locomotory signals interact on short timescales with downstream mechanosensory related signals to modulate  mechanosensory processing. 

We previously developed a high-throughput closed-loop optogenetic  approach \cite{liu_high-throughput_2022} to  interrogate  the mechanosensorimotor circuitry in \textit{C. elegans}. Here we use this method to  measure  the animal's behavior in response to over 39,000 stimulus events. From these measurements, we identified a putative circuit by which inhibitory signals from turning-associated neurons disrupt mechanosensory processing  and modulates the likelihood of a  reversal depending on the animal's behavior.

\section*{Results}
\begin{figure}[!htbp]
	\begin{center}
		\includegraphics[width=0.95\textwidth]{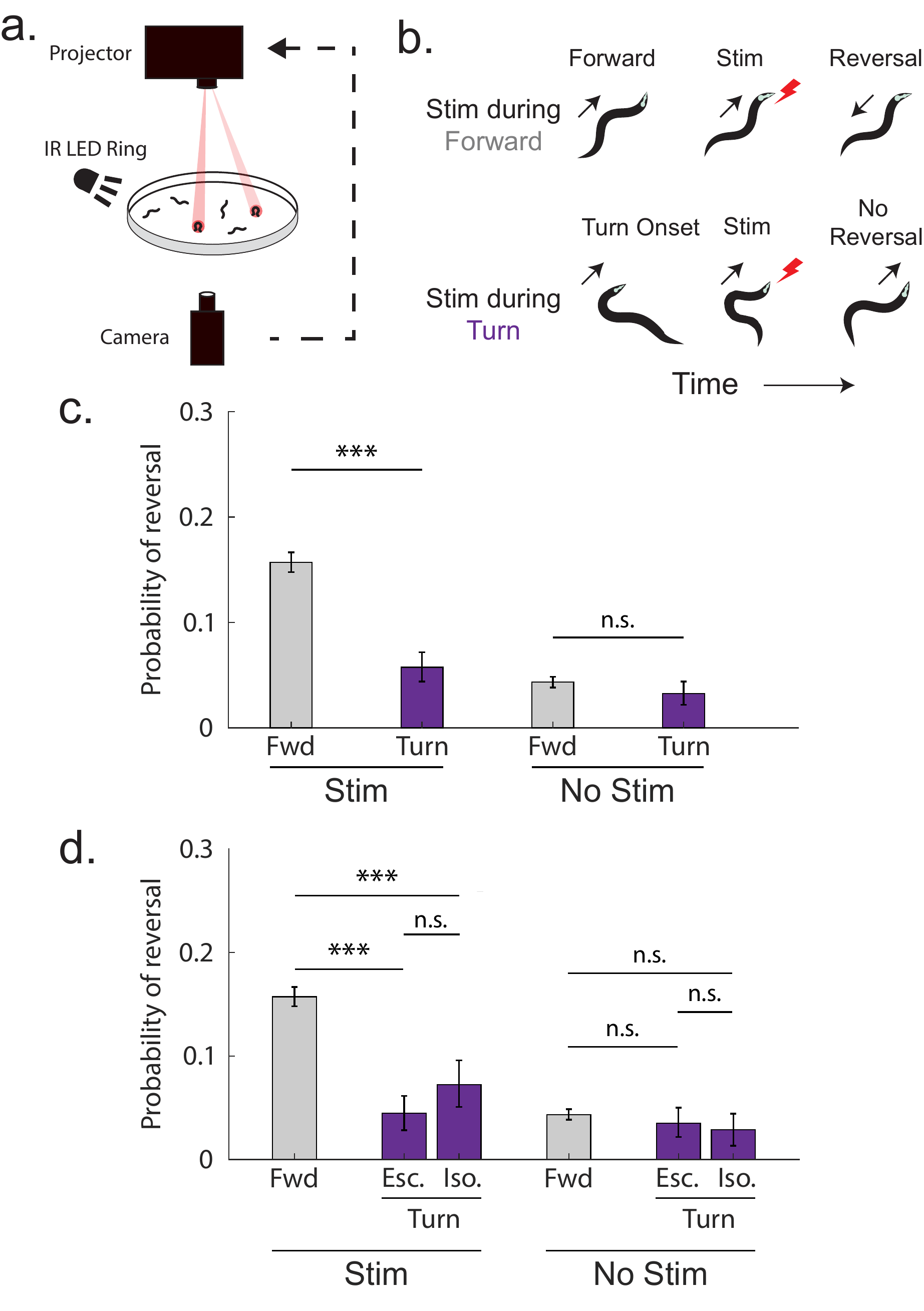}
		\caption{{\bf  Turns decrease the likelihood of mechanosensory-evoked reversals.} a) Closed-loop optogenetic stimulation is delivered to animals as they crawl based on their current behavior. b) Optogenetic stimulation is delivered to gentle-touch mechanosensory neurons in worms that are  either moving  forward (top row) or turning (bottom row).  c) The probability of a reversal is shown in response to stimulation during  forward movement or turn. Responses are also shown for a low-light no-stimulation control. This figure only is a reanalysis of recordings from  \cite{liu_high-throughput_2022}. The number of stimulation events, from left to right: 6,002, 1,114, 5,996, and 1,050. d.) The probability of reversal in response to stimulation during turning is shown broken down  further by turn subtype: escape-like turns ``Esc'' and isolated turns ``Iso.'' $N=$6,002, 602, 512, 5,996, 599 and 451 stim events, from left to right. The number of assay plates for forward and turn context are 29 and 47 respectively. The 95 percent confidence intervals for population proportions are reported.  *** indicates $p<0.001$. `n.s.' indicates $p>0.05$ via two-proportion Z-test. }

		\label{fig:mechanosensory}
	\end{center}
\end{figure}

\subsection*{Turns on their own decrease the likelihood of mechanosensory-evoked reversals}
Previously we reported that optogenetic activation of  gentle-touch mechanosensory neurons delivered during forward locomotion appeared more likely to evoke a transition to backward locomotion, called a ``reversal,''  than  activation delivered  during the onset of a turn  \cite{liu_temporal_2018}. We then developed high-throughput methods to allow us to probe this behavior with greater statistical power and concluded  that  either turning itself or possibly some other  behavior related to turning modulates  mechanosensory-evoked reversals (Fig.~\ref{fig:mechanosensory}a-c,  S1-S3 Video) \cite{liu_high-throughput_2022}. 

We sought to distinguish whether turns themselves modulated the reversals or whether it was another ancillary behavior related to turns.  Turns in our previous recordings most often occurred  immediately after backward locomotion-- part of a fixed action pattern  called  the ``escape response'' that consists of backward locomotion, a turn and then finally forward locomotion \cite{pirri_neuroethology_2012}.  By contrast, about  44\% of the turns we observed were  preceded  by only forward locomotion, what we call ``isolated'' turns. We sought to test whether isolated turns also exhibited a reduction in mechanosensory evoked responses.  

By  re-analyzing our prior   measurements  \cite{liu_high-throughput_2022}, we found that isolated turns also reduced the likelihood  of a reversal response  (Fig.~\ref{fig:mechanosensory}c,d). 
This finding suggests that turns alone are sufficient to modulate the likelihood of a mechanosensory-evoked reversal response. We therefore focused on the turn regardless of what behavior preceded it and going forward we consider both isolated- and escape-like turns together. Turning continued to modulate the likelihood of mechanosensory-evoked reversals even after animals had been stimulated multiple times and  begun showing signs of habituation Supplementary Fig.~\ref{fig:prob_of_reversal_adaptation}. And the probability of evoked reversals did not change appreciably in new experiments with modest changes of the inter-stimulus interval as shown in Supplementary Fig.~\ref{fig:inter_stim_interval_closed_loop}. In the remainder of the work we  present   results from only new experiments designed to investigate how turning modulates mechonsensory-evoked reversals.

\subsection*{Turns decrease the likelihood of interneuron-evoked reversals, except for those evoked by AVA}
\begin{figure}[p]
	\begin{center}
		\includegraphics[width=0.8\textwidth]{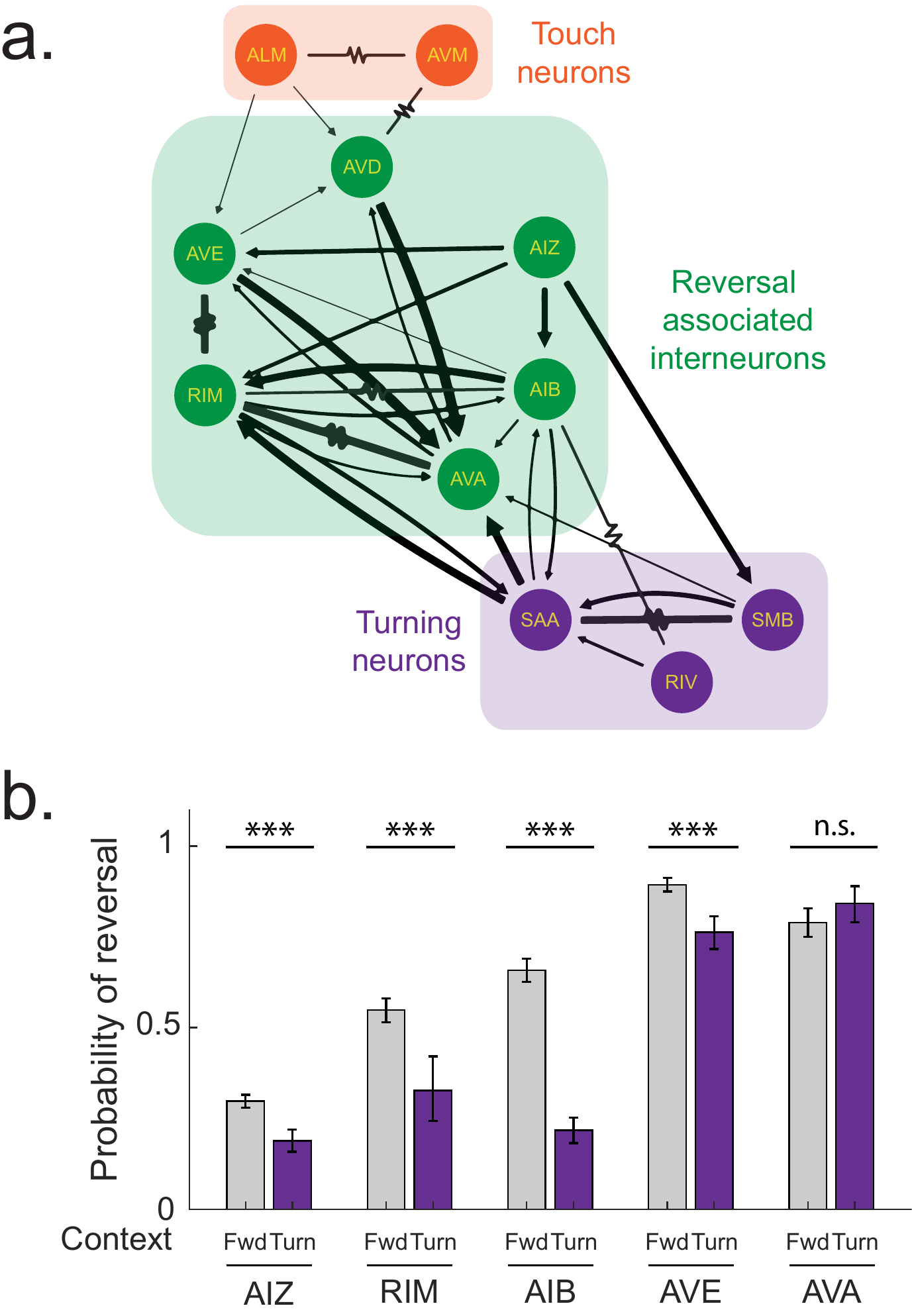}
		\caption{{\bf Turns decrease the likelihood of interneuron evoked reversals, except for AVA.} a) Anatomical connectivity showing chemical (arrows) and electrical (resistor symbol) %
		synapses among  the anterior mechanosensory neurons, downstream  interneurons, and turning associated neurons,  taken directly from nemanode.org \cite{witvliet_connectomes_2021}
		b) Probability of a reversal response is shown for 3s optogenetic stimulation to the listed neurons either during forward movement or immediately after the onset of turning. Strains are listed in Table \ref{tab:strains}. 
		Illumination was 80 $\mu$W/mm$^2$ red light  to activate   Chrimson in AVE or AVA, 300 $\mu$W/mm$^2$  blue light to activate ChR2 in RIM or AIB, and 340 $\mu$W/mm$^2$ to activate ChR2 in AIZ. Error bars indicate 95\%  confidence intervals for population proportions.  
		*** indicates $p<0.001$ via two-proportion Z-test. $p$ value for AVA stimulation group is 0.1. $N=$ [2,612, 601, 883, 107, 880, 511, 1,007, 342, 409 and 191] stimulus events, from left-to-right,
		measured  across the following number of plates: [16, 27, 12, 19, 4, 24, 8, 16, 8 and 20].}
		\label{fig:all_neurons}
	\end{center}
\end{figure}

Mechanosensory signals from the anterior gentle-touch mechanosensory neurons AVM and ALM  are thought to evoke a reversal response by traveling downstream  through a network of interneurons that are associated with  backward locomotion \cite{chalfie_neural_1985, gray_circuit_2005, chalfie_developmental_1981, wang_flexible_2020, mcclanahan_comparing_2017, mazzochette_tactile_2018, stirman_real-time_2011}. 
These include neurons AVD \cite{chalfie_neural_1985, wicks_dynamic_1996, gray_circuit_2005,  kawano_imbalancing_2011}, 
AVA \cite{shipley_simultaneous_2014, gordus_feedback_2015, piggott_neural_2011}, AIZ \cite{li_encoding_2014}, RIM \cite{clark_neural_2014, gordus_feedback_2015}, AIB \cite{gordus_feedback_2015},  AVE \cite{li_c_2020} (Fig.~\ref{fig:all_neurons}a).
Like the anterior mechanosensory neurons,  interneurons AVA, AIZ, RIM, AIB and AVE  are known to  induce reversals upon stimulation \cite{gordus_feedback_2015,li_encoding_2014, li_c_2020}. 
To better understand where this network interacts with turning, we sought to investigate whether  these interneurons' ability to evoke reversals  also depends on turning. We used a collection of transgenic strains with cell-specific or near-cell-specific promoters that drive expression of the  optogenetic proteins Chrimson or ChR2  in  each of these interneurons  (Table \ref{tab:strains}). We then  used our previously reported high-throughput closed-loop optogenetic delivery system \cite{liu_high-throughput_2022} to  stimulate the interneuron with 3 s illumination when the worm was either crawling forward or  beginning to turn. In this way we measured the animal's response to many thousands of optogenetic stimulation events.

As expected, optogenetic activation of any of the  interneurons AVA, AIZ, RIM, AIB or AVA during forward locomotion evoked reversals (Fig.~\ref{fig:all_neurons}b) compared to  the baseline probability of a spontaneous reversal  (Supplementary Fig.~\ref{fig:all_neurons_control}). 
Activating any  of the interneurons we tested, except for AVA, showed a statistically significant decrease in the probability of evoking reversals when activated during turns, compared to during forward locomotion, Fig.~\ref{fig:all_neurons}b. In other words, activation of  these interneurons showed a turning-dependent response, similar to the mechanosensory neurons.
By contrast, turning did not  significantly modulate AVA's ability to evoke reversals and the worm often aborted its turn and reversed when AVA was activated during the turn (Fig.~\ref{fig:all_neurons}b, S4 Video). 

From these perturbations  we conclude that  neurons AIZ, RIM, AIB and AVE lie  either at or upstream of the junction in which  turning signals   modulate the reversal response.  AVA, in contrast, lies in the pathway downstream of the arrival of  turning related signals. We therefore sought to investigate neural sources of this turning related signal. 

We note that for any given perturbation shown  Fig.~\ref{fig:all_neurons}, we are interested in the change of probability of  reversal  between the forward and turning contexts. We do not concern ourselves with  overall differences in reversal probability for perturbations of different  neurons because that may arise from  differences in gene expression  or differences in the efficiencies of ChR2 compared to Chrimson. Stimulation of AVD was not tested because no suitable single-cell promoter was found.

\subsection*{Turning associated neurons RIV, SMB and SAA regulate reversals}
\begin{figure}[!ht]
	\begin{center}
		\includegraphics[width=0.7\textwidth]{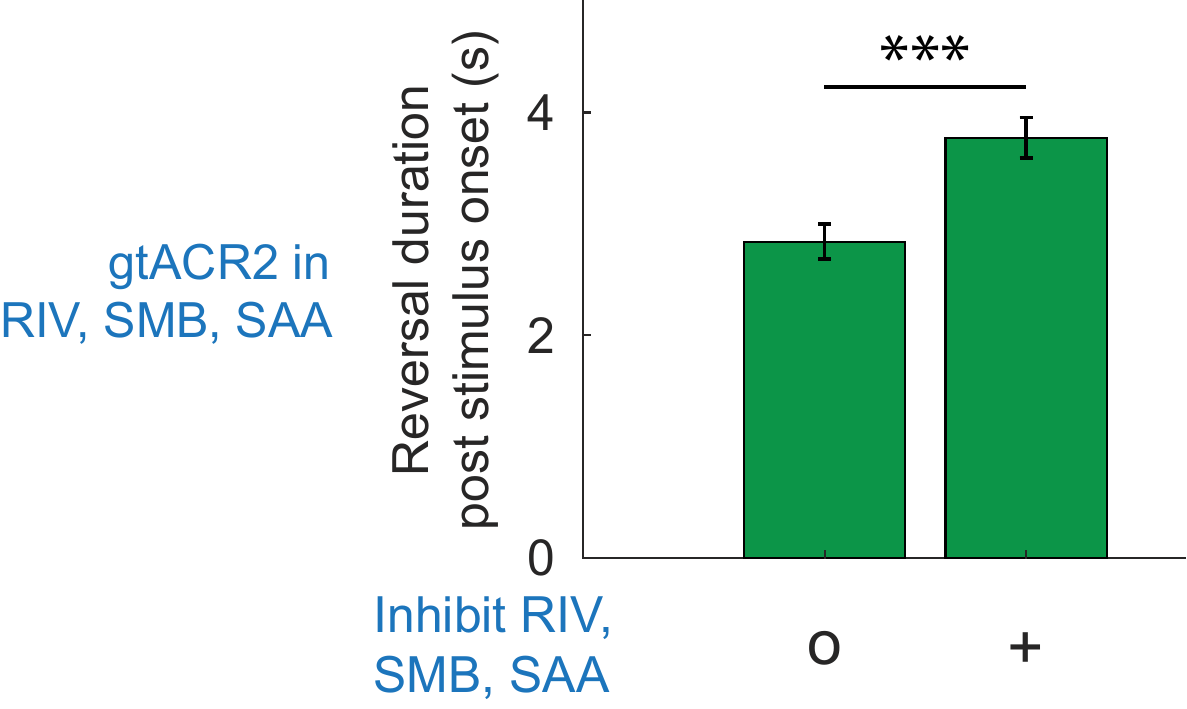}
		\caption{{\bf RIV, SMB and SAA neurons influence reversal duration.} Neurons RIV, SMB and SAA were optogenetically inhibited when worms spontaneously reversed. The  time spent going backwards is reported in a 10 s window coinciding with optogenetic inhibition upon reversal onset. Worms expressed the inhibitory opsin gtACR2 in neurons RIV, SMB, and SAA under the lim-4 promoter. Illumination intensity  of either 180 $\mu$W/mm$^2$  (`+') or  2$\mu$W/mm$^2$ (`o' control) was delivered.  Worms spent more time reversing when these neurons were inhibited than in the control. Error bars represent 95\% confidence intervals. *** indicates $p<0.001$. $N=612$ and $695$ stimulus events for `o' and `+' conditions, respectively, across 14 plates. }
		\label{fig:riv_neuron}
	\end{center}
\end{figure}

Turning in the worm  occurs either when the animal is moving forward, is paused or is transitioning from backward to forward locomotion, but not during sustained backward locomotion \cite{croll_behavoural_1975}. Neurons RIV, SMB and SAA are among those neurons associated with turning.  RIV, SMB and SAAD have increased calcium activity during turns \cite{wang_flexible_2020, kalogeropoulou_role_2018}, and ablation of RIV, SMB or SAA show defects in turning or head bending amplitude \cite{gray_circuit_2005,kalogeropoulou_role_2018}.
Wang and colleagues observed that inhibiting RIV, SMB and SAA  when the animal is backing up   prolongs the  reversal \cite{wang_flexible_2020}. They therefore  proposed that activity from turning-related neurons may inhibit  reversals.
We  independently  confirm that inhibiting RIV, SMB and SAA increases reversal duration,  Fig.~\ref{fig:riv_neuron} and Supplementary Fig.~\ref{fig:riv_mechanosensory}.  
We therefore sought to investigate whether  these turning neurons  also inhibit reversals during turns, and whether they  may explain why  mechanosensory stimulation is less likely to evoke reversals during turning.

\subsection*{Inhibiting RIV, SMB and SAA abolishes the turning dependent modulation of  mechanosensory processing}
\begin{figure}[!htbp]
	\begin{center}
		\includegraphics[width=0.7\textwidth,page=1]{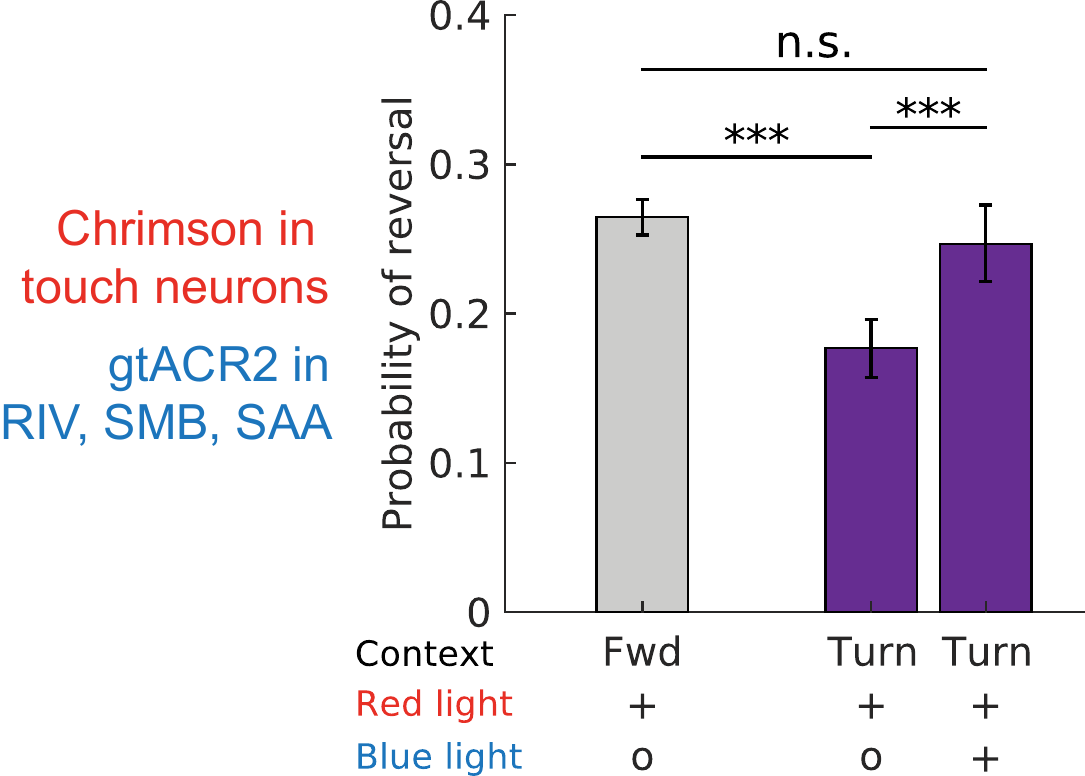}
		\caption{{\bf Optogenetic inhibition of neurons RIV, SAA and SMB during turns restore mechanosensory-evoked reversal response.} Probability of reversals  when  touch neurons are activated or when touch neurons are activated and RIV, SMB and SAA are inhibited simultaneously; during either forward movement or turn onset. Touch neurons express Chrimson and are activated  with red light.  RIV, SMB and SAA express gtACR2 and are inhibited with blue light. Strains are listed in Table \ref{tab:strains}. The 95 percent confidence intervals for population proportions are reported. *** indicates $p<0.001$, two-sample Z-test for proportions. $N=$5,381, 1,525 and 1,115 stimulation events from left to right. The number of assays from left to right bars are: $N=8, 16$, and $17$. Additional controls are shown in Fig \ref{fig:override_suppression}.}
		\label{fig:override_suppression_main}
	\end{center}
\end{figure}

We reasoned that if the turning neurons RIV, SMB and SAA inhibit reversals,  then releasing this inhibition after a turn has begun  should allow  mechanosensory stimuli delivered during the turn to evoke reversals as effectively as if they were delivered during forward locomotion. We designed an experiment to simultaneously inhibit these turning neurons while stimulating the touch neurons immediately after the onset of a turn. We expressed a blue-light inhibitory opsin, gtACR2, in the turning associated neurons RIV, SMB and SAA and a red-light activating opsin Chrimson in the gentle touch neurons.  
Inhibiting RIV, SMB and SAA  after the onset of a turn did not completely stop the  animal and it still successfully exited the turn  (see S5 Video). 
We reasoned that   ongoing RIV, SMB and SAA activity was not necessary for the completion of the turn once initiated and this therefore allowed us to  inhibit these turning associated neurons in a context in which the animal was still turning.

Activating the touch neurons by delivering red-light  immediately after the onset of a turn  was less likely to evoke a reversal than when delivered during forward locomotion, Fig.~\ref{fig:override_suppression_main}, as expected.  But when we also inhibited the RIV, SMB and SAA turning associated neurons with blue light  immediately after the turn began, the likelihood of evoking  reversals via red-light activation of the touch neurons was significantly higher and, crucially, not significantly different than for activation during forward locomotion (see S6 Video). In other words, inhibiting these turning associated neurons after turn onset  abolished the turning-dependence of the mechanosensory response.  This is consistent with a model in which signals from RIV, SMB and/or SAA   disrupt mechanosensory processing during turning. By inhibiting those neurons after the onset of a turn, we  prevent this disruption, presumably by inhibiting an inhibitory signal.

We performed additional experiments to rule out alternative explanations for why blue light illumination restored the likelihood of a mechanosensory-evoked reversal response (Supplementary Fig \ref{fig:override_suppression} and Supplementary text). For example, we find that blue light illumination when no inhibitory opsin   is present is insufficient  to restore mechanosensory evoked reversal responses during turns, suggesting that the effect is not an artifact of the blue light alone (Supplementary Fig \ref{fig:override_suppression}b).
Taken together we conclude that inhibition of the turning neurons during turns disinhibits  the mechanosensory evoked reversal response.

\subsection*{Signals from turning neurons gate mechanosensory processing }
\begin{figure}[!h]
	\begin{center}
		\includegraphics[width=0.5\textwidth]{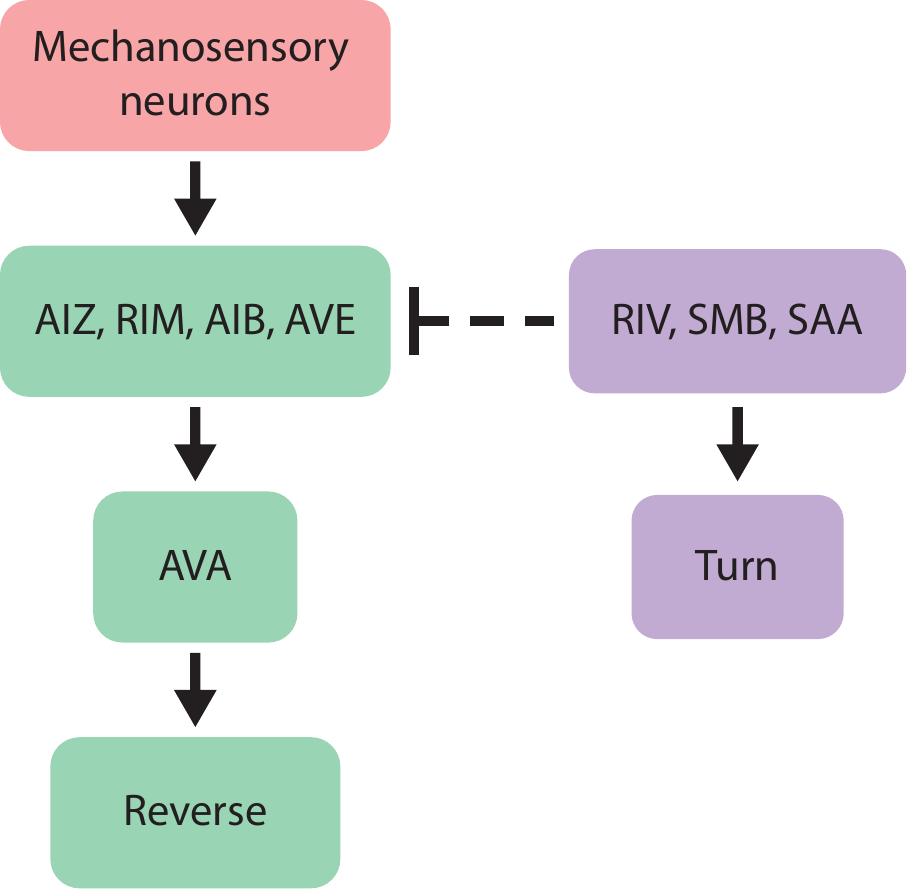}
		\caption{{\bf Putative circuit mechanism.}
		In response to gentle touch, mechanosensory neurons  propagate signals downstream through the network and reach neuron AVA to evoke a reversal. But during turning, neurons RIV, SMB, and/or SAA send inhibitory signals  that  disrupt sensory-related signals  before they reach AVA thus gating the likelihood of a reversal.}
		\label{fig:neural_mechanism}
	\end{center}
\end{figure}

Our measurements supports a model in which the turning neurons RIV, SMB and/or SAA gate mechanosensory information and prevent it from propagating further downstream to evoke a reversal, Fig.~\ref{fig:neural_mechanism}. In this model,  mechanosensory signals from the gentle-touch mechanosensory  neurons ALM and AVM propagate downstream in a feedforward manner to reversal-associated interneurons RIM, AIZ, AIB and AVE. If the animal is moving forward, the mechanosensory signals continue to propagate to AVA and evoke reversals. But if  the animal is turning, inhibitory signals originating from RIV/SMB/SAA suppress or disrupt mechanosensory-related signals within the interneurons and  prevent downstream mechanosensory-related signals from propagating to AVA. This model is consistent with our measurements and  leads us to conclude that turning-related inhibitory signals gates downstream mechanosensory processing.

\section*{Discussion and Conclusions}
Here we show that putative inhibitory signals from turning associated neurons RIV/SMB/SAA modulate mechanosensory evoked reversals downstream of the gentle touch neurons and upstream of neuron AVA. But within those constraints, where exactly might those signals combine? Neuron wiring  and gene expression data suggests that one location may be across the inhibitory synapses from SAA to AIB and RIM. SAA is cholinergic and makes chemical synapses onto AVA, AIB and RIM \cite{white_structure_1986, witvliet_connectomes_2021}. While the synapses onto AVA are functionally excitatory \cite{randi_functional_2022}, the ones onto AIB and RIM may be inhibitory because AIB expresses  the inhibitory acetylcholine receptors \textit{lgc-47}, and \textit{acc-1}; while RIM expresses inhibitory (e.g. \textit{lgc-47} \cite{taylor_molecular_2021}, and \textit{acc-1} \cite{huo_hierarchical_2023}) and excitatory (e.g. \textit{acr-3})  acetylcholine receptors.
We note that  AIB and RIM both synapse onto AVA, therefore SAA mediated inhibition of AIB and RIM may decrease overall excitation to AVA, broadly consistent with our cartoon model in Fig.~\ref{fig:neural_mechanism}.

Wang and colleagues had previously predicted that turning circuitry may inhibit reversal circuitry  \cite{wang_flexible_2020}. Now in contemporaneous work from the same group,  Huo and colleagues show that activation of SAA/RIV/SMB terminates reversals and inhibits RIM when RIM is already in an active state, likely through an ACC-1 acetylcholine-gated chloride channel  \cite{huo_hierarchical_2023}. 

Our findings  are consistent with the mechanism proposed  in \cite{huo_hierarchical_2023} in which SAA blocks reversals by inhibiting  RIM. More broadly our findings reinforce a longstanding hypothesis that different motor programs in the worm  inhibit one another, as was previously proposed for forward and reverse locomotion \cite{zheng_neuronal_1999}.

In our model, AVA performs a role similar to that of a ``decision neuron''  with respect to reversals \cite{buetfering_behaviorally_2022}.   This is  consistent with our previous observation that AVA's calcium activity more closely reflects the animal's decision to reverse, and is less reflective of the  strength of the stimulus (e.g. AVA's activity does not reflect how many touch neurons are activated) \cite{shipley_simultaneous_2014}. The simple model we describe assumes feed-forward propagation of signals from ALM and AVM through the downstream network to AVA and omits recurrent connections among the neurons in between. Future investigations are needed to explore additional contributions from recurrence in the network, and of the role of AVD, for which we lacked a cell-specific promoter.

More broadly, we show that motor related signals are directly influencing neural activity in areas that  contain a mix of sensory and motor information.  This is reminiscent of saccadic suppression in vision \cite{bremmer_neural_2009, binda_vision_2018, turner_visual_2022} and corollary discharge \cite{crapse_corollary_2008, ji_corollary_2021, riedl_tyraminergic_2022} in which motor related activity  modulates or impinges upon sensory representations. Our findings  add to a growing body of evidence suggesting that behavior information is necessary for sensory processing, and this may explain why behavior-related neural activity patterns are seen across the brain in mice, fly and worms, including in nominally sensory areas \cite{hallinen_decoding_2021, stringer_spontaneous_2019, musall_single-trial_2019, atanas_brain-wide_2022}.

Because turning events are infrequent, spontaneous and brief,  they are rare compared to the time the animal spends moving forward or backwards.  But  obtaining  sufficient statistical power to probe sensory processing during turns required  hundreds of observations per condition. In total we  measured over 40,000 behavior responses to stimulation, including more than 16,000  during turns.  This investigation was therefore only made feasible by leveraging the recent high-throughput methods we presented in \cite{liu_high-throughput_2022} that use computer-vision and targeted illumination to   track many worms in parallel and to automatically deliver stimuli triggered upon the animal's turns.

\section*{Materials and methods}
\subsection*{Strains}
Strains used in this work are listed  in Table \ref{table:methods:strains}. In each strain light-gated ion channels have been expressed to either excite or inhibit specific neurons. We expressed excitatory opsin Chrimson in the six gentle touch neurons using the \textit{mec-4} promoter. Promoters \textit {ser-2, tdc-1, npr-9, opt-3, rig-3} are used to express excitatory opsin in neurons AIZ, RIM, AIB, AVE, and AVA respectively. To express gtACR2 in RIV, SMB, and SAA, we used the \textit{lim-4} promoter and performed integration using a mini-SOG approach. We injected into CZ20310 worms, followed by a blue light treatment (450nm, M450LP1, Thorlabs) for 30 minutes as described in \cite{noma_rapid_2018}. Before conducting experiments, we outcrossed integrated worms with the wild type N2 strains for at least six generations to generate AML496. AML496 worms were then crossed into AML67 worms to create AML499 strain. Our transgenic strains include a mix of WT and \textit{lite-1} mutant backgrounds. We measured no systematic difference in locomotion or to endogenous  blue-light response in these two backgrounds for the light levels and conditions used here Supplementary Fig.~\ref{fig:health_of_worms}.

\begin{table}
\begin{adjustwidth}{-2.35in}{0in}
\centering
\caption{ \label{tab:strains}
{\bf Strains  used.}} 
\begin{tabular}{|c|p{1.7cm}|p{1.7cm}|p{6.2cm}|p{3.1cm}|p{1.4cm}|p{0.3cm}|}
\hline
Strain name & Target neuron expression & additional expression & Genotype & Figure & Ref   \\\hline
AML67 & ALML, ALMR, AVM, PLML, PLMR, PVM  & & wtfIs46[mec-4P::Chrimson::SL2::mCherry::unc-54 40ng/ul] & Fig \ref{fig:mechanosensory}c,d, Fig \ref{fig:override_suppression}b, Fig \ref{fig:inter_stim_interval_closed_loop}, and Fig \ref{fig:health_of_worms} & \cite{liu_temporal_2018}\\ \hline
TQ3301 & AIZ & & xuIs198[Pser-2(2)::frt::ChR2::YFP,Podr-2(2b)::flp, Punc-122::YFP]; lite-1(xu7) & Fig \ref{fig:all_neurons}b, Fig \ref{fig:all_neurons_control}, and Fig \ref{fig:health_of_worms} & \cite{li_encoding_2014}\\ \hline
QW910 & RIM & & zfIs9[Ptdc-1::ChR2::GFP, lin-15+]; lite-1(ce314) & Fig \ref{fig:all_neurons}b, Fig \ref{fig:all_neurons_control}, and Fig \ref{fig:health_of_worms} & \cite{clark_neural_2014}\\ \hline
QW1097  & AIB & & zfIs112[Pnpr-9::ChR2::GFP, lin15+]; lite-1(ce314) & Fig \ref{fig:all_neurons}b, Fig \ref{fig:all_neurons_control}, and Fig \ref{fig:health_of_worms} & \cite{clark_neural_2014}\\ \hline
Not provided  & AVE & & opt-3::Chrimson & Fig \ref{fig:all_neurons}b, Fig \ref{fig:all_neurons_control}, and Fig \ref{fig:health_of_worms} & \cite{li_c_2020}\\ \hline
AML17  & AVA & I1, I4, M4, and NSM & wtfIs2[rig-3::Chrimson::SL2::mCherry] & Fig \ref{fig:all_neurons}b, Fig \ref{fig:all_neurons_control}, and Fig \ref{fig:health_of_worms} &  \cite{shipley_simultaneous_2014}\\ \hline
AML496  & RIV, SMB, SAA & & wtfIs465 [lim-4P::gtACR2::SL2::eGFP::unc-54 80ng/ul + unc-122::RFP 50ng/ul] & Fig \ref{fig:riv_neuron}, Fig \ref{fig:override_suppression}c, and Fig \ref{fig:health_of_worms} & This work\\ \hline
AML499  & RIV, SMB, SAA; ALML, ALMR, AVM, PLML, PLMR, PVM  & & wtfIs46[mec-4P::Chrimson::SL2::mCherry::unc-54 40ng/ul]; wtfIs465 [lim-4P::gtACR2::SL2::eGFP::unc-54 80ng/ul + unc-122::RFP 50ng/ul] & Fig \ref{fig:override_suppression_main}, Fig \ref{fig:riv_mechanosensory}, Fig \ref{fig:override_suppression}a, and Fig \ref{fig:health_of_worms} & This work\\ \hline
N2  & -  & & - & Fig \ref{fig:health_of_worms} & \\ \hline
KG1180  & -  & & lite-1(ce314) & Fig \ref{fig:health_of_worms} & \cite{edwards_novel_2008}  \\\hline

\end{tabular}
\label{table:methods:strains}
\end{adjustwidth}
\end{table}

\begin{table}[!h]
\centering
\includegraphics[width=1\linewidth]{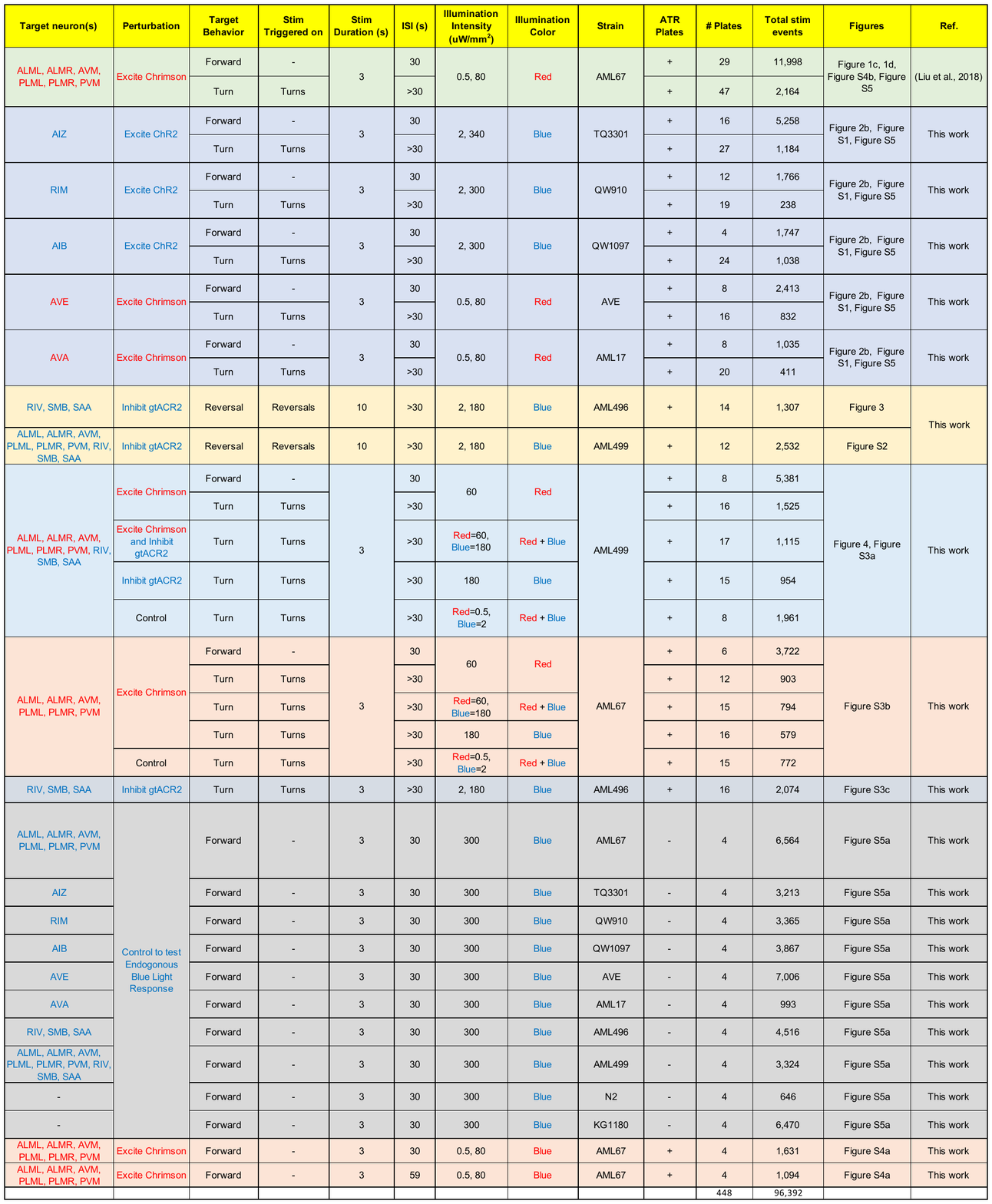}
\caption{List of optogenetic measurements performed during behavior.}
\label{tab:dataset}
\end{table}

\subsection*{Nematode handling}
All worm strains were maintained at 20 C, on regular NGM media plates seeded with \textit{E. coli} (OP-50) as food source. Experiments were performed on young adult animals. To obtain young adults, worms were bleached three days prior to the experiments. %
Bleached eggs were  washed and centrifuged  in M9 (0.8rcf for two mins)  three times. Bleached eggs were  suspended in M9 and stored in a shaker overnight. The following  morning hatched L1 larvae were centrifuged and transferred to freshly seeded plates consisting of 1 ml of 0.5 mM all-trans-retinal mixed with OP50 and stored in the dark at 20 C until young adulthood. 

For experiments, young adult worms were washed in M9 and  transferred to an empty agarose plate for experiments. Excess M9 solution was absorbed with a kimwipe as described in \cite{liu_high-throughput_2022,liu_temporal_2018}.

\subsection*{Behavior Analysis}
Computer-vision based behavior analysis was used to identify when the animal is moving forward, when it is undergoing a reversal or when it is turning. The closed loop latency from detecting a turn to delivering an optogenetic stimulation is 167 ms \cite{liu_high-throughput_2022}. Analysis was performed as reported previously using two different sets of algorithms, one for real time applications and the other retrospectively in post-processing \cite{liu_high-throughput_2022}.  All figures in this work reflect behavior classifications  from the off-line retrospective analysis. 

Briefly, animals are segmented and a centerline is detected.  Additional logic is used to find centerlines even when the animal touches itself \cite{liu_temporal_2018}. The animal's center of mass velocity is also computed.   Behavior classification is first performed by  classifying pose dynamics in a behavior map \cite{liu_temporal_2018, berman_mapping_2014} and then refined by inspecting the animal's ellipse ratio and center of mass velocity to catch any omitted turns, or instances when the behavior mapper fails to classify.  Compared to our previous recent work \cite{liu_high-throughput_2022} we changed two  parameters to be more conservative in classifying  animals as turning or reversing. Specifically, to be classified as turning we now require that the binary image of the animal have an ellipse ratio of 3.1, compared to 3.6 previously. Similarly, to be classified as a reversal, the animal must now achieve a center of mass velocity of  -0.11 mm/s, instead of -0.1 mm/s, during the 3 s optogenetic stimulus window. These changes were minor and were implemented to catch rare events that previously had escaped classification.

For experiments probing reversal duration, we report the time the animal spent going backwards in a 10 s window, coinciding with optogenetic inhibition. 10 s was chosen because it was a compromise between the 12 s used in  \cite{wang_flexible_2020} and the shorter stimuli that we typically use \cite{liu_high-throughput_2022}.  So for example, if after stimulus onset the animal continued moved backwards for 3 s, then paused for 1 s, and moved backwards for 2 s more, we report a ``reversal duration'' of 5 s.

\subsection*{Optogenetic activation and inhibition}
In this work we seek to deliver optogenetic illumination  specifically when the animal is either moving forward, or turning, or reversing. We conduct different  sets of experiments for each of these three conditions, using different sets of animals for each experiment. In all cases we use a projector-based illumination system that tracks many individuals on a plate full of animals, segments them in real time, and  addresses each animal individually to shine light on them, as described previously \cite{liu_high-throughput_2022}. All experiments are performed on  plates containing approximately 30 to 40 animals.

To measure the animal's response to optogenetic activation or inhibition delivered during the onset of turns, our system waited until it detected that an animal was beginning to turn, and  then delivered a stimulus automatically. In post-processing we retrospectively evaluated whether the turn was valid at time of stimulus onset, and only included those stimuli events that met our more stringent criteria, as described in \cite{liu_high-throughput_2022}. 

To measure the animal's response to optogenetic perturbations during forward locomotion, we optogenetically illuminated all tracked animals on the plate every 30 s, in open loop. In post-processing we then only considered those animals that were moving forward at the time of illumination. The worms in the open loop assays were stimulated every 30 seconds. However, in the closed loop experiments, the worms were stimulated when turns were detected. As a result, the worms received optogentic stimulus less frequently, shown in Fig.~\ref{fig:inter_stim_interval_closed_loop}b.  There was no statistical significance difference in the probability of evoked reversal for stimuli delivered during forward locomotion in these two conditions Fig.~\ref{fig:health_of_worms}b.

To measure the animal's response to optogenetic inhibition during reversals, our system waited until it detected that an animal had been reversing for 1 second, and then delivered the illumination. As before, we retrospectively confirmed that the animal was reversing before including it for further analysis. 

Illumination color, intensity and  duration are listed in Table \ref{tab:dataset}.

\subsection*{Statistical Analysis}
To reject the null hypothesis that two empirically observed probabilities are the same, we use a two-proportion Z-test \cite{noauthor_733_nodate}.  Error bars report 95\% confidence interval calculated via a bootstrap procedure.

\subsection*{Data availability}
Computer-readable files showing processed tracked behavior trajectories and stimulus events for all experiments are publicly posted at  \url{https://doi.org/10.6084/m9.figshare.21699668}.

\subsection*{Code availability}
All analysis code used in this manuscript are available at \url{https://github.com/leiferlab/analysis-code-kumar-2022.git}.

\subsection*{Strains and plasmid availability}
All genetic strains and plasmids generated as part of this manuscript are being made available through Caenorhabditis Genetics Center (CGC) and Addgene respectively.

\section*{Acknowledgments}
We thank  Zhaoyu Li (Queensland Brain Institute), Shawn Xu (University of Michigan) and Mark Alkema (University of Massachusetts Worcester) for  strains. We thank Matthew Creamer for helpful discussions. This work used computing resources from the Princeton Institute for Computational Science and Engineering.  
Research reported in this work was supported  by the Simons Foundation under award  SCGB \#543003 to AML;  and  by the National Science Foundation, through an NSF CAREER Award to AML (IOS-1845137) and through the Center for the Physics of Biological Function (PHY-1734030). Strains from this work are being distributed by the CGC, which is funded by the NIH Office of Research Infrastructure Programs (P40 OD010440). The content is solely the responsibility of the authors and does not  represent the official views of any funding agency.

\nolinenumbers

\bibliographystyle{plos2015}

\newpage
\section*{Supplementary Text and Figures}
\renewcommand{\thefigure}{S\arabic{figure}}
\renewcommand{\figurename}{Supplementary Fig}
\setcounter{figure}{0}

\begin{figure}[!htbp]
	\begin{center}
	\includegraphics[width=0.7\textwidth]{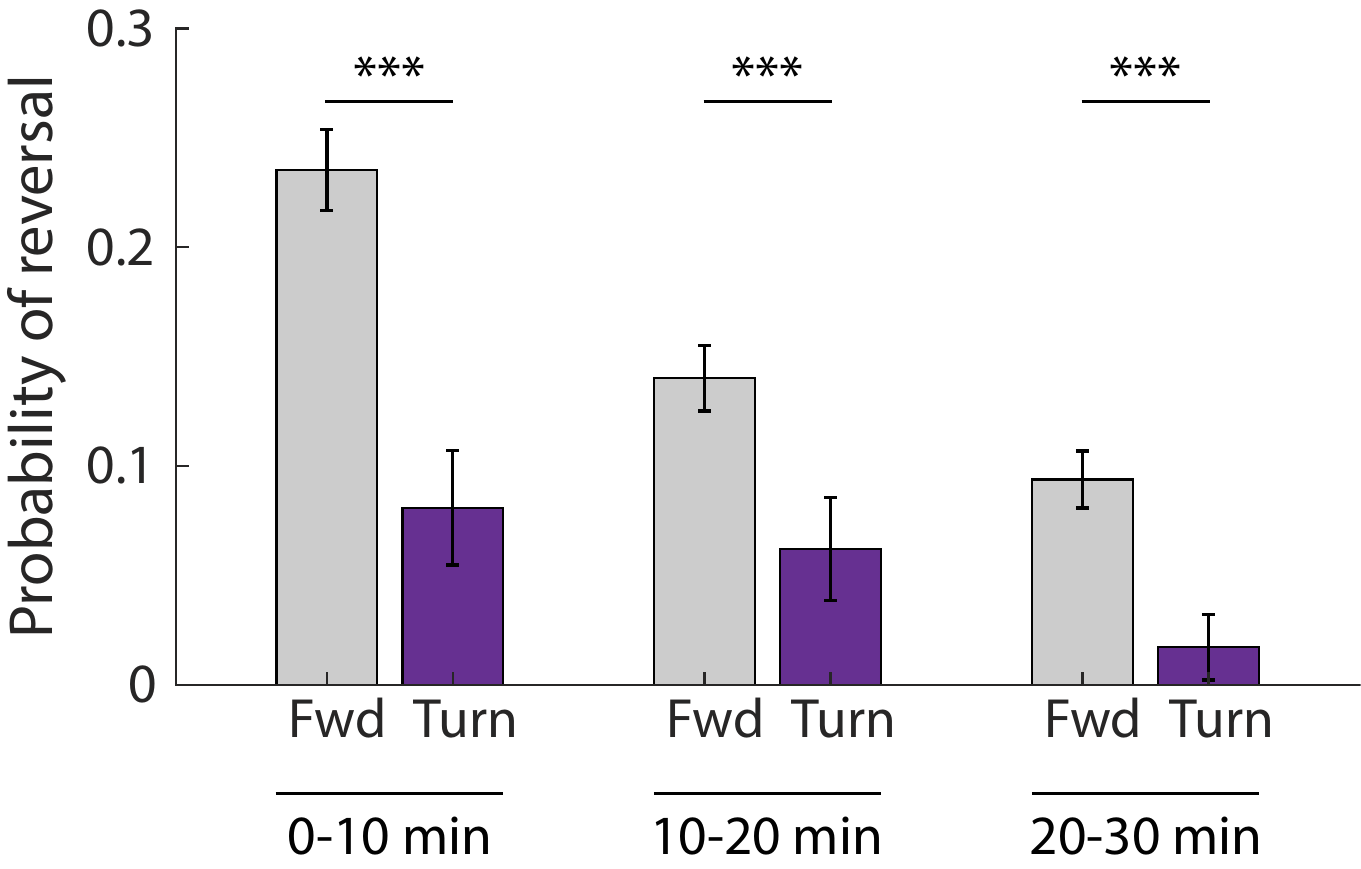}
    \caption{{\bf Probability of reversing in response to stimuli delivered during turns  is consistently lower than for stimuli delivered during forward locomotion throughout the duration of the 30 min assay.} Probability of evoked reversal in response to optogenetic stimulation to gentle touch mechanosensory neurons (\textit{pmec-4}::Chrimson) is calculated for three different portions   of the 30 min experiment. Habituation is visible, but the relative difference in reversal probability persists.  Error bars show 95 percent confidence intervals of the population proportions.  *** indicates $p<0.001$ via two sample z-test, $N=$2,006, 420, 2,077, 403, 1919, and 291 stimulation events from left to right. The number of assay plates for forward and turn context are $N=$ 29 and 47 respectively. This figure is a reanalysis of measurements presented in \cite{liu_high-throughput_2022}. }
    \label{fig:prob_of_reversal_adaptation}
    \end{center}
\end{figure}

\begin{figure}[!htbp]
	\begin{center}
		\includegraphics[width=0.8\textwidth]{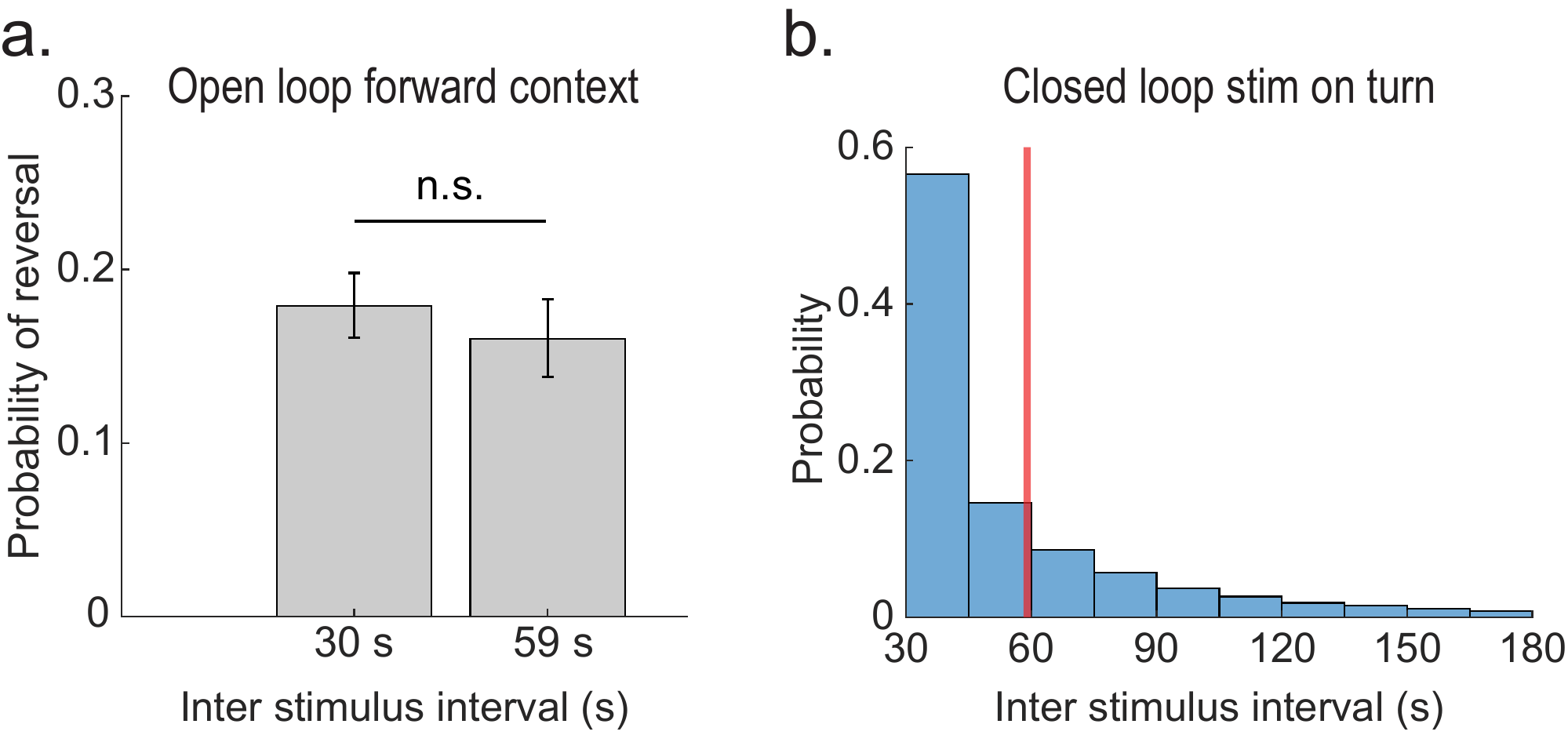}
		\caption{{\bf Probability of reversal is similar for two different inter-stimulus intervals.} a)  Animals expressing Chrimson in their gentle touch mechanosensory neurons  were optogenetically  stimulated in open loop every  30 s or  59 s.   Only responses to  stimuli delivered during forward locomotion are included. These are new experiments not previously reported. $N=$1,631 and 1,094 stim events.  Error bars show 95 percent confidence intervals of the population proportions.  b) 59 s (vertical red bar) is the mean inter stimulus interval (ISI) experienced by worms in the closed-loop turn-triggered stimulus experiments previously presented in \cite{liu_high-throughput_2022}. The ISI is not constant because it depends on when the worm turns.  The  distribution of the ISI experienced by worms during those experiments  Fig.~\ref{fig:mechanosensory} is shown in blue.}
		\label{fig:inter_stim_interval_closed_loop}
	\end{center}
\end{figure}

\begin{figure}[!htbp]
	\begin{center}
	\includegraphics[width=0.7\textwidth]{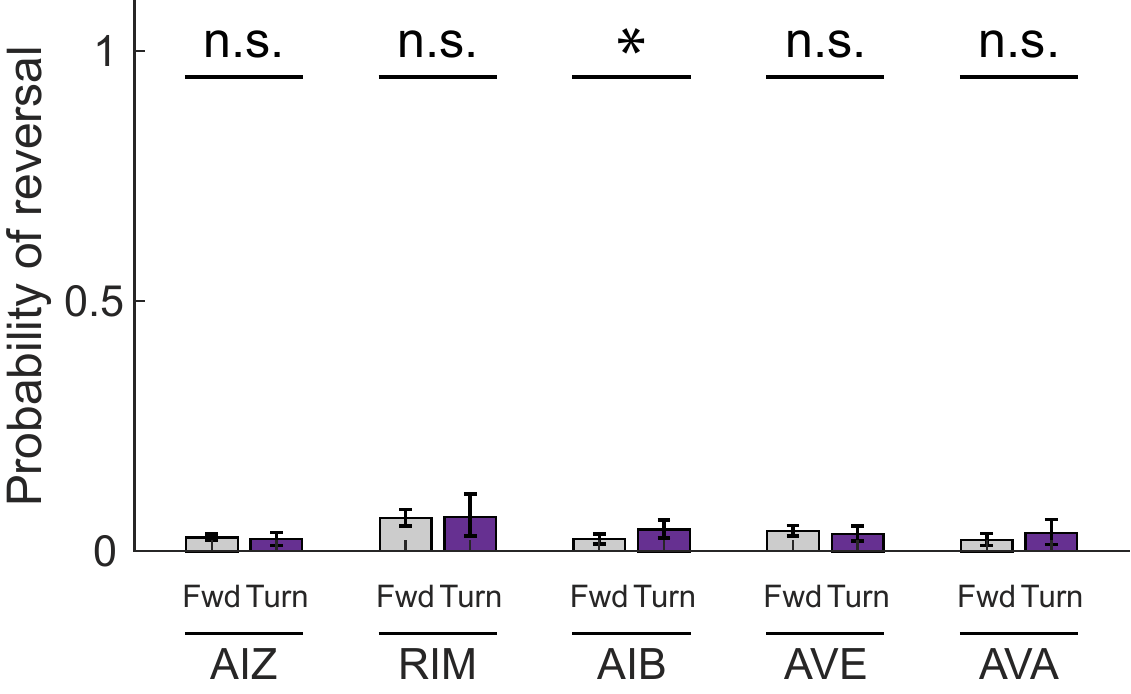}
    \caption{{\bf Baseline reversal probabilities measured via low-light (control) illumination.} a) Baseline reversal probabilities for each strain in each condition are measured  by shining a low-intensity control stimulus. Three seconds of only 0.5 $\mu$W/mm$^2$ of red light illumination  (neuron AVE and AVA) or 2 $\mu$W/mm$^2$ of blue light illumination (neuron AIZ, RIM, and AIB). The 95 percent confidence intervals for population proportions are reported. Two sample Z-test was used to calculate significance. $p$ value for AIZ, RIM, AIB, AVE, and AVA stimulation group is 0.596, 0.936, 0.045, 0.565, 0.262 respectively. The number of stimulus events for each condition (from left-most bar to right-most bar) are: 2,646, 583, 883, 131, 867, 527, 1,406, 490, 626, 220.  The number of assay plates for forward and turn context for neurons from left to right are 16, 27, 12, 19, 4, 24, 8, 16, 8, 20.}
    \label{fig:all_neurons_control}
    \end{center}
\end{figure}

\begin{figure}[!htbp]
	\begin{center}
	\includegraphics[width=0.7\textwidth]{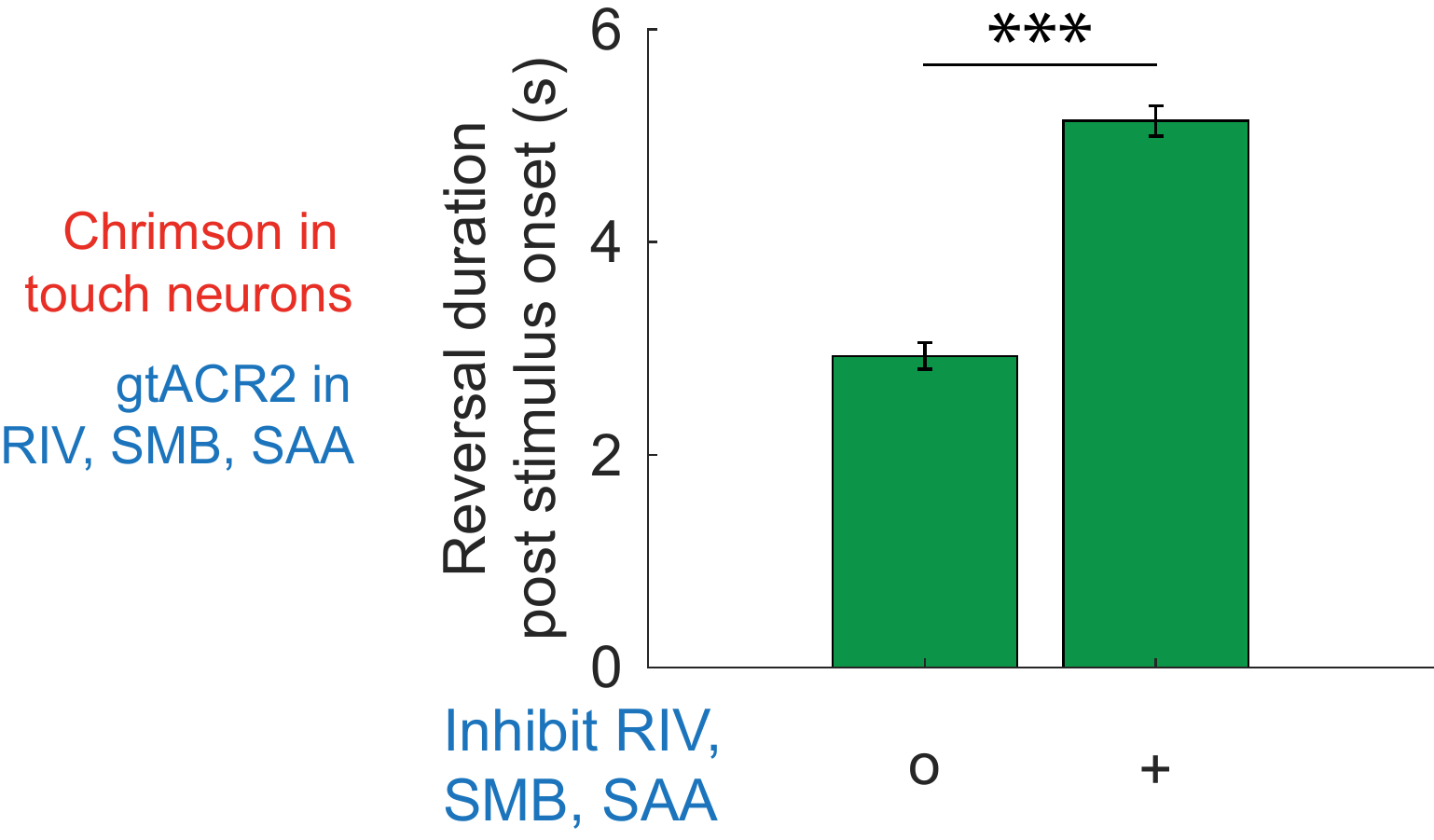}
    \caption{{\bf Inhibition of RIV, SMB and SAA prolong reversals, in a second transgenic background.} Same experiment as in Fig.~\ref{fig:riv_neuron}, but in a transgenic background that also expresses Chrimson in the mechanosensory neurons. Results are consistent with Fig.~\ref{fig:riv_neuron}. Worm spent more time reversing when the RIV, SMB, and SAA neurons were inhibited compared to when a control stimulus intensity was used. Error bars represent 95\% confidence intervals. *** indicates $p<0.001$. The number of stimulus events for mock and experimental conditions are 1,168, 1,364 respectively.  The number of assays was $N=14$.}
		\label{fig:riv_mechanosensory}
	\end{center}
\end{figure}

\begin{figure}[!htbp]
	\begin{center}
		\includegraphics[width=0.8\textwidth]{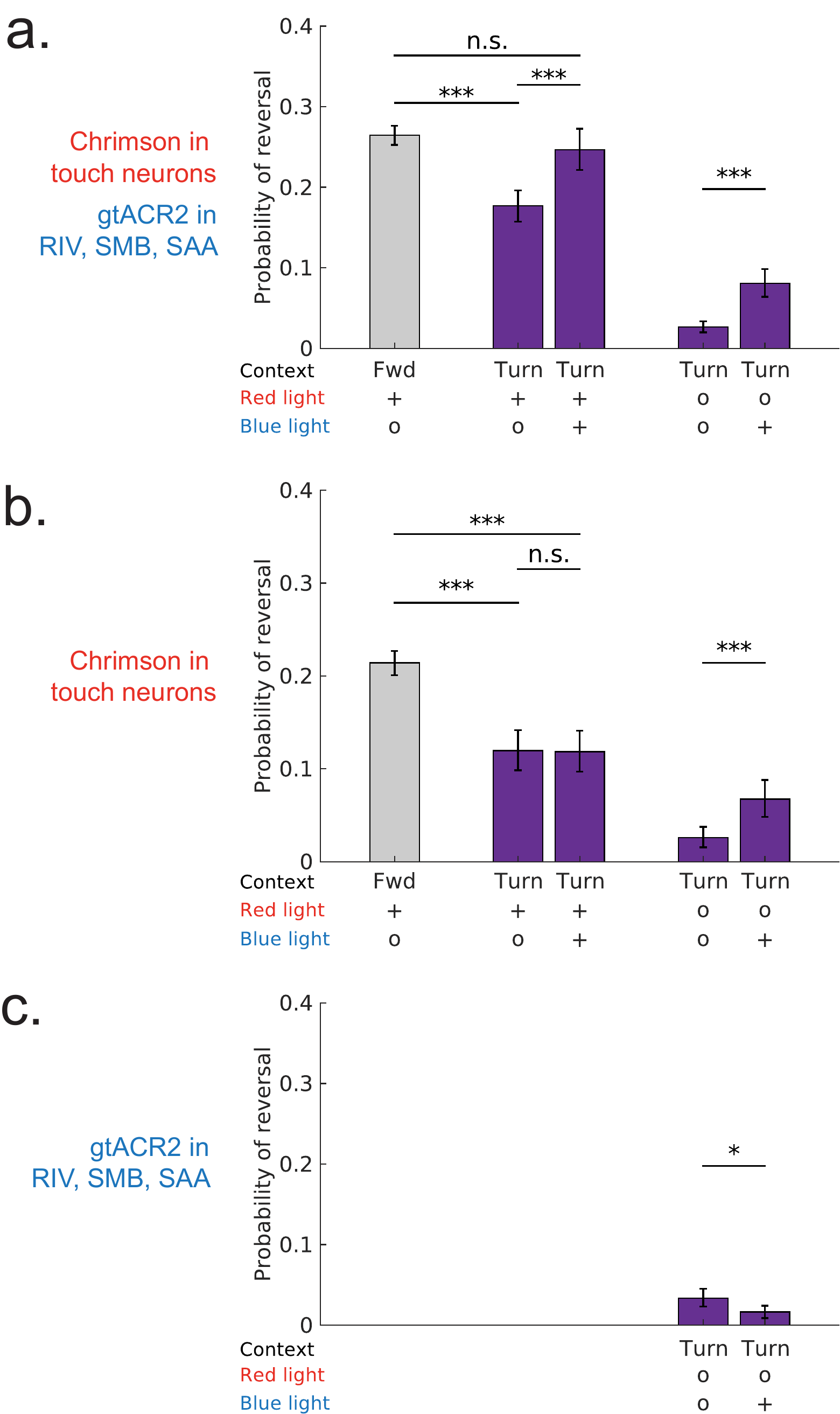}
		\caption{{\bf Additional control experiments show that blue-light alone cannot restore  mechanosensory-evoked reversal response.} a) Probability of reversals  when either touch neurons are activated, or RIV, SMB and SAA are inhibited, or both simultaneously; during either forward movement or turn onset. First three bars are same as in Figure \ref{fig:override_suppression}. Touch neurons express Chrimson and are activated  with red light.  RIV, SMB and SAA expressing gtACR2  are  inhibited with blue light. Strains are listed in Table \ref{tab:strains}. The 95 percent confidence intervals for population proportions are reported. *** indicates $p<0.001$, two-sample z-test for proportions. $N=$5,381, 1,525, 1,115, 954 and 1,961 stim events, from left to right. The number of assays from left to right bars are: $N=8, 16, 17, 15$, and $8$. b) Same experiments were repeated in a strain that  expressed Chrimson in the gentle-touch mechanosensory neurons, but no inhibitory opsins.   $N=$3,722, 903, 794, 579 and 772 stim events. The number of assays from left to right bars are: $N=6, 12, 15, 16$, and $15$. c) Same experiments are shown  for animals that only express inhibitory opsin gtACR2 in RIV, SMB and SAA, but no Chrimson. $N=$1,041 and 1,033 stim events. The number of assay is: $N=16$.}
		\label{fig:override_suppression}
	\end{center}
\end{figure}

\begin{figure}[!htbp]
	\begin{center}
		\includegraphics[width=0.8\textwidth]{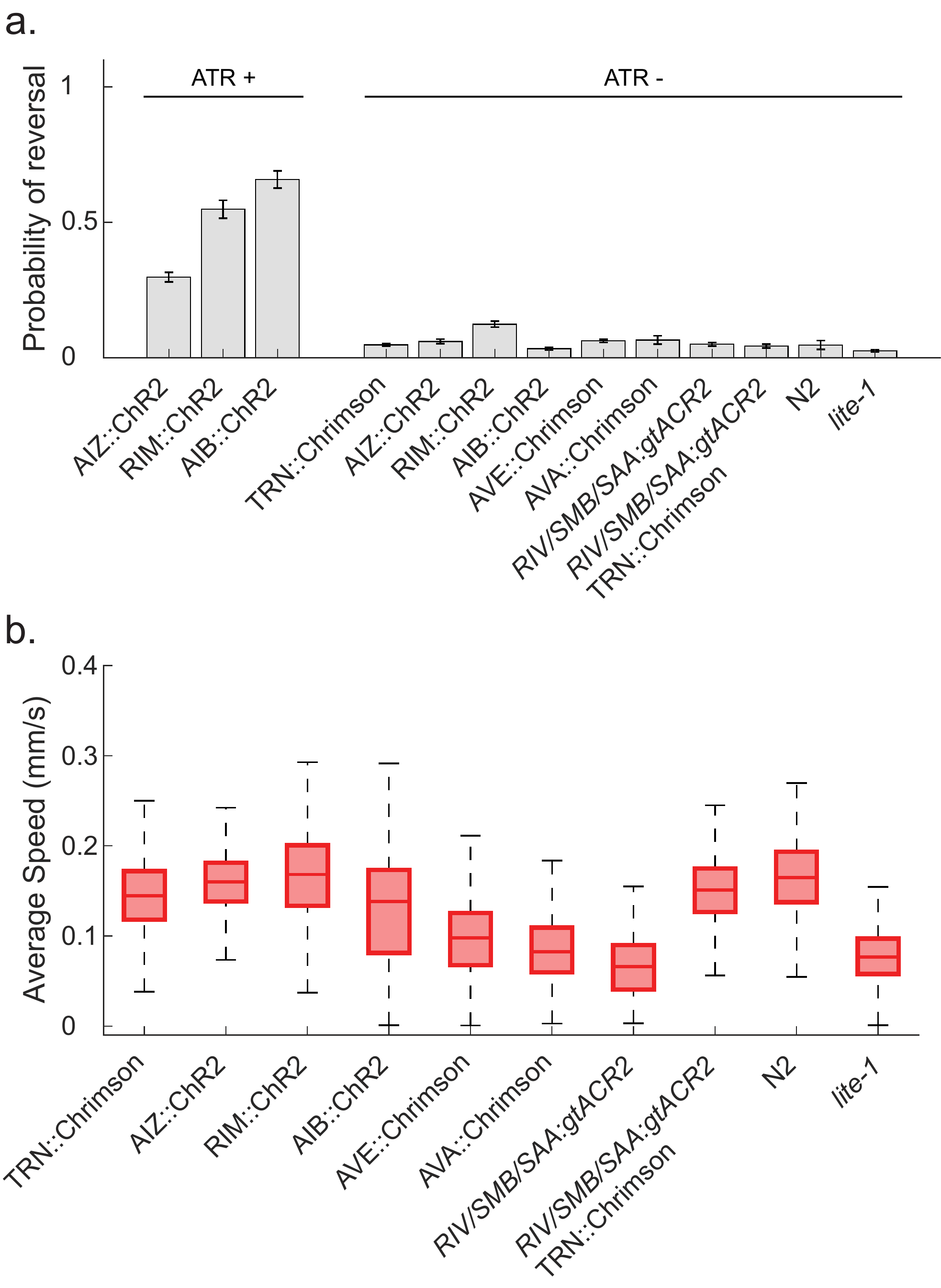}
		\caption{{\bf Blue-light sensitivity and baseline locomotion activity of strains used.}
	a) Blue light-evoked reversal probability is shown for different strains with and without the all-trans retinal (ATR) co-factor needed for optogenetic proteins.  Here we show those strains that express ChR2 (left, N=2612, 883, 880 from left to right) in the ATR+ condition (same as Figure \ref{fig:all_neurons}b), and we also measure all strains in  an  ATR- condition including Chrimson strains (right, N=6564, 3213, 3365, 3867, 7006, 993, 4516, 3324, 646, 6470 from left to right). Error bars show 95 percent confidence intervals for population proportions. We  include a \textit{lite-1} mutant and wild-type N2 for comparison because our transgenic strains include a mix of both wild-type and \textit{lite-1} backgrounds. b) Average speed of each  strains used in this work are shown N=1706, 1654, 983, 1099, 1251, 837, 564, 1065, 654, 1952 from left to right. }
		\label{fig:health_of_worms}
	\end{center}
\end{figure}

\newpage
\subsubsection*{Additional control experiments show that blue-light alone cannot restore mechanosensory evoked reversals}
We sought to rule out alternative explanations for why shining blue light during turns may cause an increase in reversals.  It is known that the nominally red-light sensitive Chrimson  can also be mildly activated by blue light \cite{klapoetke_independent_2014}. We therefore we tested whether the increase in responsiveness to the mechanosensory stimulus  was the result of blue-light activation of Chrimson expressed in the touch neurons. 

Consistent with mild blue-light activation of Chrimson, shining only blue light on animals expressing Chrimson in the touch neurons (Supplementary Fig.~\ref{fig:override_suppression}a,b far right bar) but not on animals expressing only gtACR2 in the turning neurons (Supplementary Fig.~\ref{fig:override_suppression}c far right bar) caused a significant increase in the probability of reversals compared to no stimulation (second to right bar). However, this mild blue-light activation of Chrimson  is insufficient to explain the increase in reversal probability we observed when inhibiting turning neurons via gtACR2.  Even when shining both blue and red light on animals that lack the inhibitory opsin gtACR2, but do contain Chrimson in their touch neurons,  we still observed a large and significant reduction in the likelihood of reversing in response to stimuli delivered during turns compared to delivered during forward locomotion (Supplementary Fig.~\ref{fig:override_suppression}b middle bar, compared to far left bar). This suggests that it is the inhibition of  neurons RIV, SMB and SAA that abolishes  the turning-dependence of mechanosensory processing and not mild blue-light activation of the touch neurons. Further consistent with this view,  adding  blue light to red light in those animals that lack the inhibitory opsin does not  significantly increase  the probability of reversing (Supplementary Fig.~\ref{fig:override_suppression}b second compared to third bar). 

A simple and fully consistent explanation is that our red light illumination   strongly activates the touch neurons and that any additional blue light contributes only very modest additional activation to the touch neurons, and not enough to explain the increase we see when inhibiting RIV/SMB/SAA. Moreover, this modest additional blue-light activation of the touch neurons is only  significant in control experiments without  any red light. Taken together we conclude that inhibition of the turning neurons, and not mild blue-light activation of Chrimson,  is responsible for abolishing the turning dependence of the mechanosensory response.

\newpage

\paragraph*{S1 Video}
\label{vid:V1_Video}
Example showing behavior of a population of animals during an experiment from \cite{liu_high-throughput_2022}. Middle 24 s of a 30 minute recording is shown. Optogenetic stimulation is delivered in closed loop when  turning of an individual animal is detected. Each yellow numbered `x' represents a tracked animal, with its track shown in yellow. Inset at top left shows detailed movements of  worm number 213, denoted by a green square. The head of the worm is represented by green dot. A centerline is drawn through the worm's body and is shown in green. The dynamic circular pattern of green and white spots in the center of the video is a visual timestamp system projected onto the plate that is used for synchronizing the timing of video analysis, as described in \cite{liu_high-throughput_2022}. \href{https://vimeo.com/823005066?share=copy}{https://vimeo.com/823005066}

\paragraph*{S2 Video}
\label{vid:V2_Video}
Example of a worm reversing in response to optogenetic stimulation of its gentle-touch mechanosensory neurons delivered during forward locomotion. Recording is from \cite{liu_high-throughput_2022}. Animals express Chrimson in gentle touch mechanosensory neurons (strain name: AML67). Stimulus was delivered in open-loop.  Green dot denotes the animal's head. Green line denotes its centerline. Yellow line shows the trajectory of  a point midway along the animal's centerline over the past 10 seconds. Red indicates area illuminated by red light. 
\href{https://vimeo.com/823006874}{https://vimeo.com/823006874}

\paragraph*{S3 Video}
\label{vid:V3_Video}
Example of a worm receiving optogenetic stimulation of its gentle touch mechanosensory neurons during the onset of a turn. Recording is from \cite{liu_high-throughput_2022}. Animals express Chrimson in gentle touch mechanosensory neurons (strain name: AML67). This worm does not reverse in response to stimulation. Stimuli was triggered in closed-loop by the animal's turn. Green dot denotes the animal's head. Green line denotes its centerline.  Yellow line shows the trajectory of  a point midway along the animal's centerline over the past 10 seconds. Red indicates area illuminated by red light. 
\href{https://vimeo.com/823007966}{https://vimeo.com/823007966}

\paragraph*{S4 Video}
\label{vid:V4_Video}
Example of a worm aborting a turn and reversing when neuron AVA was activated following the onset of the turn. Animals express Chrimson in neuron AVA (strain name: AML17). Stimulation was delivered upon the onset of a turn in closed-loop. Green dot denotes the animal's head. Green line denotes its centerline. Yellow line shows the trajectory of  a point midway along the animal's centerline over the past 10 seconds. Red indicates area illuminated by red light. 
\href{https://vimeo.com/823008477?share=copy}{https://vimeo.com/823008477}

\paragraph*{S5 Video}
\label{vid:V5_Video}
Example of a worm completing a turn during inhibition of neurons RIV, SMB and SAA. Animals express  the inhibitory opsin gtACR2 in these neurons (strain name: AML496). Stimulation was delivered upon the onset of a turn in closed-loop.  Green dot denotes the animal's head. Green line denotes its centerline.  Yellow line shows the trajectory of  a point midway along the animal's centerline over the past 10 seconds. Blue indicates area illuminated by blue light. 
\href{https://vimeo.com/823011657?share=copy}{https://vimeo.com/823011657}

\paragraph*{S6 Video}
\label{vid:V6_Video}
Example of a worm aborting a turn and reversing when   neurons RIV, SMB and SAA are inhibited and the gentle-touch  mechanosenseory neurons are activated (strain name: AML499). Animals express  the inhibitory opsin gtACR2 in RIV, SMB and SAA and the excitatory opsin Chrimson in the gentle -touch mechanosensory neurons. Blue and red light illumination was delivered upon the onset of a turn in closed-loop.  Green dot denotes the animal's head. Green line denotes its centerline.  Yellow line shows the trajectory of  a point midway along the animal's centerline over the past 10 seconds. Purple  indicates area illuminated by light. 
\href{https://vimeo.com/823011676?share=copy}{https://vimeo.com/823011676}

\end{document}